\documentclass[aps,prd,twocolumn,nofootinbib,showpacs]{revtex4}

\usepackage{graphicx}

\begin{document}

\title{Application of the Principle of Maximum Conformality to the Hadroproduction of  the Higgs Boson at the LHC}

\author{Sheng-Quan Wang$^{1}$}
\email[email:]{sqwang@cqu.edu.cn}

\author{Xing-Gang Wu$^{2}$}
\email[email:]{wuxg@cqu.edu.cn}

\author{Stanley J. Brodsky$^{3}$}
\email[email:]{sjbth@slac.stanford.edu}

\author{Matin Mojaza$^{4}$}
\email[email:]{mojaza@nordita.org}

\address{$^1$School of Science, Guizhou Minzu University, Guiyang 550025, P.R. China}
\address{$^2$Department of Physics, Chongqing University, Chongqing 401331, P.R. China}
\address{$^3$SLAC National Accelerator Laboratory, Stanford University, Stanford, California 94039, USA}
\address{$^4$Nordita, KTH Royal Institute of Technology and Stockholm University, Roslagstullsbacken 23, SE-10691 Stockholm, Sweden}

\date{\today}

\begin{abstract}

We present improved pQCD predictions for Higgs boson hadroproduction at the Large Hadronic Collider (LHC) by applying the Principle of Maximum Conformality (PMC), a procedure which resums the pQCD series using the renormalization group (RG), thereby eliminating the dependence of the predictions on the choice of the renormalization scheme while minimizing sensitivity to the initial choice of the renormalization scale. In previous pQCD predictions for Higgs boson hadroproduction, it has been conventional to assume that the renormalization scale $\mu_r$ of the QCD coupling $\alpha_s(\mu_r)$ is the Higgs mass, and then to vary this choice over the range $1/2 m_H < \mu_r < 2 m_H $ in order to estimate the theory uncertainty. However, this error estimate is only sensitive to the non-conformal $\beta$ terms in the pQCD series, and thus it fails to correctly estimate the theory uncertainty in cases where pQCD series has large higher order contributions, as is the case for Higgs boson hadroproduction. Furthermore, this \mbox{\it ad hoc} choice of scale and range gives pQCD predictions which depend on the renormalization scheme being used, in contradiction to basic RG principles. In contrast, after applying the PMC, we obtain next-to-next-to-leading order RG resummed pQCD predictions for Higgs boson hadroproduction which are renormalization-scheme independent and have minimal sensitivity to the choice of the initial renormalization scale. Taking $m_H=125$ GeV, the PMC predictions for the $p p \to H X$ Higgs inclusive hadroproduction cross-sections for various LHC center-of-mass energies are: $\sigma_{\rm Incl}|_{\rm 7\,TeV} = 21.21^{+1.36}_{-1.32}$ pb, $\sigma_{\rm Incl}|_{\rm 8\,TeV} = 27.37^{+1.65}_{-1.59}$ pb, and $\sigma_{\rm Incl}|_{\rm 13\,TeV} = 65.72^{+3.46}_{-3.01}$ pb, respectively. We also predict the fiducial cross section $\sigma_{\rm fid}(pp\to H\to\gamma\gamma)$: $\sigma_{\rm fid}|_{\rm 7\,TeV}=30.1^{+2.3}_{-2.2}$ fb, $\sigma_{\rm fid}|_{\rm 8\,TeV}=38.3^{+2.9}_{-2.8}$ fb, and $\sigma_{\rm fid}|_{\rm 13\,TeV}=85.8^{+5.7}_{-5.3}$ fb. The error limits in these predictions include the small residual high-order renormalization-scale dependence, plus the uncertainty from the factorization-scale. The PMC predictions show better agreement with the ATLAS measurements than the LHC-XS predictions which are based on conventional renormalization scale-setting.


\end{abstract}

\maketitle

\section{Introduction}

The Higgs boson predicted by the Standard Model (SM) was discovered by ATLAS and CMS Collaborations at the Run I stage of the Large Hadron Collider (LHC)~\cite{Aad:2012tfa, Chatrchyan:2012xdj}. This remarkable discovery initiated a new era of precision studies of Higgs phenomenology. The specific properties of the Higgs boson are now being probed in LHC Run II. The comparison of SM predictions with the new data will test the electroweak symmetry breaking mechanism and probe possible new physics beyond the SM, as discussed by the LHC Higgs Cross Section Working Group (the LHC-XS group)~\cite{Heinemeyer:2013tqa}. The details of the hadronic production of the Higgs plays an important role for understanding this fundamental phenomenology. Experimentally, the first measurements of the total and differential cross sections for the inclusive $p p \to H X$ production channel, followed by the decays $H\to\gamma\gamma$ or $H\to ZZ^{*}\to 4l$, have been reported by the ATLAS Collaboration at proton-proton CM collision energies $\sqrt{s}=7$ TeV, $8$ TeV and 13 TeV~\cite{Aad:2015lha, TOTCS:ATLAS}. Theoretically, the Higgs hadroproduction cross section has been calculated up to next-to-next-to-leading order (NNLO)~\cite{Harlander:2002wh, Anastasiou:2002yz, Ravindran:2003um, deFlorian:2012yg}. A state-of-the-art, next-to-next-to-next-leading order (NNNLO) analysis of the dominant gluon-fusion production channel has recently been performed in Ref.\cite{Anastasiou:2015ema}. These calculations provide the basis for highly precise tests of pQCD predictions.

A key requirement of the renormalization group (RG) is that the prediction for a physical observable must be independent of the choice of renormalization scheme as well as the initial choice of the renormalization scale. The higher-order pQCD predictions for Higgs hadroproduction are currently based on conventional scale-setting~\cite{Harlander:2002wh, Anastasiou:2002yz, Ravindran:2003um, deFlorian:2012yg, Anastasiou:2015ema}, where one assumes the Higgs mass ($m_{H}$) itself is the renormalization scale and then varies it over an arbitrary range -- typically $[m_H/2,2m_H]$ -- in order to ascertain the scale uncertainty. However, the conventional scale-setting procedure leads to a dependence on the renormalization scheme and scale which cannot be repaired by a high fixed-order calculation. A higher order calculation could soften this scale-dependence to a certain degree, but it cannot solve the problem. Furthermore, the convergence of the resulting pQCD series is questionable due to the presence of divergent renormalon terms of order $n ! \, \alpha^n_s \beta_0^n$ which emerge at higher-orders. The estimate of the theory uncertainty which is obtained by simply varying the renormalization scale is also unreliable, since it only accounts for contributions from higher-order ``non-conformal" terms, while ignoring the contributions from the ``conformal" ($\beta=0$) terms which appear at the same order. As pointed out in Ref.\cite{Forte:2013mda}, these problems become even worse for Higgs hadroproduction: If one uses the conventional error estimate, the calculated higher-order predictions are systematically outside of the error bars predicted from the lower-order cross-sections, thus showing the importance of the conformal terms at the same perturbative order.

Large ``$K$-factors" and other normalization uncertainties are often observed for many high-energy processes, indicating poor pQCD convergence; however, one cannot decide whether the problem is an intrinsic property requiring a resummed perturbative series or a signal indicating the wrong choice of scales. For example, the cross section for $e^+ e^- \to b \bar b$ near threshold involves gluonic virtuality both of order $s$ and order ${v_{\rm rel}}^2 s$, where $v_{\rm rel}$ is the heavy quark relative velocity and $s$ is the $e^+ e^-$ center-of-mass energy squared~\cite{Brodsky:1995ds}. It is generally expected that one can soften the scheme and scale dependence by including higher-and-higher order contributions; optimistically, this procedure could be relevant for predictions for a global quantity such as a total cross-section or a total decay width because of the cancelation of scale errors at progressively higher orders; however, one does not have any certainty that one has reliable predictions for cross-section or decay width at any finite perturbative order; two examples are presented in Refs.\cite{Wang:2015lna, Zeng:2015gha}. Moreover, this procedure will clearly fail for pQCD predictions for differential observables where multiple renormalization scales appear, reflecting differing gluon virtualities at different orders and at different phase-space points.

It should be emphasized, that as in quantum electrodynamics (QED), the relevant renormalization scale is typically different at each order, reflecting the different virtualities of the relevant amplitudes. For example, the ``increasing-decreasing" behavior observed by the D0 collaboration for $A_{\rm FB}$ as the $t \bar t$ invariant mass is increased~\cite{Abazov:2014cca} reflects the fact that the physical scales (and the effective number of quark flavors $n_f$) for the one-gluon and two-gluon $s$-channel skeleton amplitudes which contribute to the front-back $t \bar t$ asymmetry are quite different; this asymmetry is not evident using conventional single-scale renormalization scale-setting, even by a NNLO QCD calculation -- one predicts a monotonically increasing behavior~\cite{Czakon:2016ckf}. Thus a renormalization scale-setting approach which can take into account multiple physical scales is essential for precise pQCD collider predictions and for reliable comparisons with the experimental measurements~\cite{Wu:2013ei, Wu:2015rga}.

The recently developed Principle of Maximum Conformality (PMC)~\cite{Brodsky:2011ta, Brodsky:2011ig, Mojaza:2012mf, Brodsky:2013vpa} provides an unambiguous way to eliminate the conventional renormalization scheme-and-scale ambiguities. The PMC has a solid theoretical foundation and satisfies the essential property of RG-invariance~\cite{Brodsky:2012ms, Wu:2014iba}. The PMC provides the underlying principle for the well-known Brodsky-Lepage-Mackenzie (BLM) method~\cite{Brodsky:1982gc}~\footnote{A generalization of BLM to higher orders in a renormalization scale- and scheme- invariant manner in large-$\beta_0$ limit by using the Pade approximant has been presented in Ref.\cite{Brodsky:1997vq}.}; it generalizes the BLM procedure by shifting all $\beta$ terms into the scale of the running coupling at all orders, and it reduces to the standard scale-setting procedure of Gell-Mann and Low (GM-L)~\cite{GellMann:1954fq} in the $N_C \to 0$ QED Abelian limit~\cite{Brodsky:1997jk}. As in QED, separate renormalization scales and effective numbers of quark flavors $n_f$ appear for each skeleton graph, correctly reflecting their differing virtualities.

The running behavior of a QCD coupling in any renormalization scheme is governed by the $\beta$-function entering its RG-equation. The $\beta$ terms entering a pQCD series can then be used to determine the optimized `physical' renormalized scales of the process. For example one can generalize the conventional $\overline{\rm MS}$ dimensional regularization procedure by simply subtracting $\ln 4\pi -\gamma_E -\delta$ instead of $\ln 4\pi -\gamma_E$ from the ultraviolet (UV) divergent terms, thus defining the $R_\delta$ scheme~\cite{Mojaza:2012mf, Brodsky:2013vpa}. The coefficients of the $\delta$ terms of the pQCD prediction in the $R_\delta$ scheme unambiguously determines the occurrence and pattern of the $\beta$ contributions at every order. The $\beta$ terms are then systematically eliminated by the PMC by shifting the argument of the relevant running coupling for each skeleton graph, thus setting its renormalization scale. After these shifts, the resulting pQCD series matches that of the corresponding ``conformal" series with $\beta=0$. Then, after normalizing the value of the coupling $\alpha_s(Q)$ from a measurement at a single momentum transfer $Q$, the resulting predictions are scheme-independent at each order.

The scheme-independence of the PMC predictions for observables is clear for the $R_\delta$-schemes, since the scheme-related $\delta$-dependent terms are eliminated by PMC scale-setting procedure at each order~\cite{Mojaza:2012mf, Brodsky:2013vpa}. The PMC predictions for physical observables are thus independent of the choice of renormalization scheme, a key requirement of renormalization group invariance. Scheme-independence is also ensured by the commensurate scale relations (CSRs) relating different schemes or observables to each other~\cite{Brodsky:1994eh}~\footnote{An analysis of the CSRs up to high perturbative orders has been given in Refs.\cite{Brodsky:2013vpa}; a detailed PMC analysis of the residual scheme- dependence of CSRs relating various observables is in preparation.}.

In contrast to conventional scale-setting, which assumes a single renormalization scale for all orders, the PMC scales for each order are generally different due to the different $\beta$-patterns which emerge at each order. As a byproduct, after applying the PMC, the divergent renormalon series such as $\sum n !\, \alpha_s^n \beta_0^n$ does not appear; the convergence of pQCD is thus greatly improved.

The PMC has been successfully applied to a number of higher-order processes; a recent review is given in Ref.\cite{Wu:2015rga}. For example, it provides a comprehensive, self-consistent pQCD explanation for both the top-pair total production cross-section and the top-pair forward-backward asymmetry measured by the LHC and Tevatron collaborations~\cite{Brodsky:2012sz, Brodsky:2012rj, Brodsky:2012ik, Wang:2014sua, Wang:2015lna}. In this paper we will investigate whether more precise and more reliable pQCD predictions for the Higgs boson hadroproduction can be achieved by applying the PMC.

The remainder of this paper is organized as follows. In Sec.~\ref{sect2}, we present the detailed technology needed for applying the PMC to Higgs boson hadroproduction. In Sec.~\ref{sect3}, we present numerical results specific to the LHC. Sec.~\ref{sect4} is reserved for a summary.

\section{PMC scale-setting for the hadroproduction of the Higgs boson}
\label{sect2}

\begin{figure}[htb]
\centering
\includegraphics[width=0.45\textwidth]{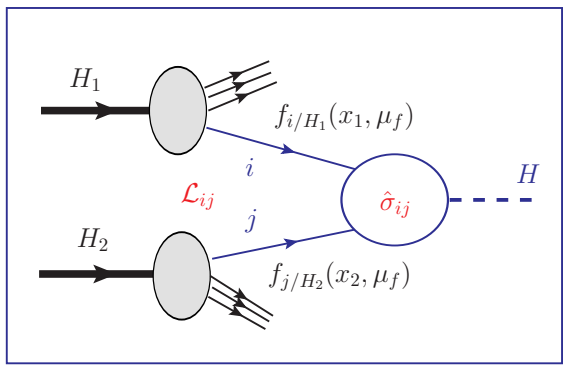}
\caption{Diagrammatic illustration of Higgs boson production at hadron colliders, computed from the convolution of partonic cross-sections $\hat \sigma_{ij}$ with the corresponding parton luminosities ${\cal L}_{ij}$. } \label{figure}
\end{figure}

The cross-section for the production of the Higgs boson in proton-proton collisions, as illustrated in Fig.(\ref{figure}), can be treated as the convolution of the hard-scattering partonic cross-section $\hat \sigma_{ij}$ with the corresponding parton luminosities ${\cal L}_{ij}$, i.e.
\begin{equation}
\sigma_{H_1 H_2 \to {H X}} = \sum_{i,j} \int\limits_{m^2_{H}}^{S}\, ds \,\, {\cal L}_{ij}(s, S, \mu_f) \hat \sigma_{ij}(s,M,R) , \label{basic}
\end{equation}
with the parton luminosity
\begin{displaymath}
{\cal L}_{ij} = {1\over S} \int\limits_s^S {d\hat{s}\over \hat{s}} f_{i/H_1}\left(x_1,\mu_f\right) f_{j/H_2}\left(x_2,\mu_f\right),
\end{displaymath}
where the summation indices $i,j$ run over all possible parton flavors in proton $H_1$ or $H_2$, $x_1= {\hat{s} / S}$ and $x_2= {s / \hat{s}}$. Here $S$ denotes the hadronic center-of-mass energy squared, and $s=x_1 x_2 S$ is the subprocess center-of-mass energy squared. Each subprocess cross-section $\hat \sigma_{ij}$ depends on the renormalization scale $\mu_r$, and the parton luminosities depend on the factorization scale $\mu_f$. We also introduce the useful ratios $M=\mu_f^2/m_H^2$ and $R=\mu_r^2/\mu_f^2$, where $m_H$ is the Higgs boson mass. The parton distribution functions (PDF) underlying the parton luminosities $f_{i/H_{\alpha}}(x_\alpha,\mu_f)$ ($\alpha=1$ or $2$) describe the probability of finding a parton of type $i$ with light-front momentum fraction $x={k^+\over P^+}$ between $x_\alpha$ and $x_{\alpha} +dx_{\alpha}$ in the proton $H_{\alpha}$. Furthermore, by setting $s=m_H^2 (S/m_H^2)^{y_1}$ and $\hat{s}=s(S/s)^{y_2}$, we can transform the two-dimensional integration over $s$ and $\hat{s}$ into an integration over two variables $y_{1,2}\in[0,1]$. These integrals can be performed numerically using the VEGAS program~\cite{Lepage:1977sw}.

The pattern of $\beta$ terms entering a pQCD series consistent with the RG can be identified at each perturbative order by using the $R_\delta$ method~\cite{Mojaza:2012mf}; however, for some processes this identification can be accomplished at low orders by noting the occurrence of the $n_f$ terms which are associated with $\beta_0$ and $\beta_1$. This is the procedure used by the BLM method. The resulting PMC renormalization scales, obtained by shifting the arguments of the running couplings to absorb the $\beta$ terms at each relevant order, reflect the gluonic virtualities. Analytic expressions with explicit $n_f$-dependence up to NNLO level can be found in Refs.\cite{Anastasiou:2002yz, Ravindran:2003um}, which are calculated using the $\overline{\rm MS}$-scheme and can be directly adopted for our PMC analysis. More explicitly, the partonic cross-section $\hat{\sigma}_{ij}$ up to NNLO level can be written as
\begin{widetext}
\begin{eqnarray}
\hat\sigma_{ij}(s,M,R)&=&\frac{\pi}{576 v^2} \left[ \eta_{ij}^{(0)}(s,M,R) a_s^{2}(\mu_r)
 + \eta_{ij}^{(1)}(s,M,R) a_s^{3}(\mu_r) + \eta_{ij}^{(2)}(s,M,R) a_s^{4}(\mu_r)+{\cal O}(a_s^5) \right],
\label{eq:partonicexpansion}
\end{eqnarray}
\end{widetext}
where $v\simeq 246$ GeV is the Higgs boson vacuum expectation value, $a_s = \alpha_s/4\pi$ with $\alpha_s$ being the strong coupling constant. The perturbative coefficients $\eta_{ij}^{(0)}$, $\eta_{ij}^{(1)}$ and $\eta_{ij}^{(2)}$ can be written in $n_f$-series as
\begin{eqnarray}
\eta_{ij}^{(0)}(s,M,R)&=&c^{ij}_{1,0}(s,M,R), \label{eq:cij1} \\
\eta_{ij}^{(1)}(s,M,R)&=&c^{ij}_{2,0}(s,M,R)+c^{ij}_{2,1}(s,M,R)n_f, \label{eq:cij2} \\
\eta_{ij}^{(2)}(s,M,R)&=&c^{ij}_{3,0}(s,M,R)+c^{ij}_{3,1}(s,M,R)n_f \nonumber\\
&& +c^{ij}_{3,2}(s,M,R)n^2_f.
\label{eq:cij3}
\end{eqnarray}
At the LO level,
\begin{eqnarray}
\eta_{ij}^{(0)}(s,M,R)=4^2\delta_{ig} \delta_{jg}\delta(1-m_H^2/s), \label{pQCDcoe1}
\end{eqnarray}
which shows that only the $(gg)$-channel is non-zero at this order. At the NLO level,
\begin{eqnarray}
&& \eta_{gg}^{(1)}(s,M,R) \neq 0 \;,\; \eta_{qg}^{(1)}(s,M,R)=\eta_{\bar{q}g}^{(1)}(s,M,R) \neq 0 \; ,\; \nonumber \\
&& \eta_{q\bar{q}}^{(1)}(s,M,R) \neq 0 \;,\; \eta_{qq'}^{(1)}(s,M,R) = \eta_{\bar{q}\bar{q}'}^{(1)}(s,M,R) = 0 , \label{pQCDcoe2}
\end{eqnarray}
where $q'$ may or may not be equal to $q$. At the NNLO level, all perturbative coefficients are non-zero; i.e.,
\begin{eqnarray}
&& \eta_{ij}^{(2)}(s,M,R) \neq 0 . \label{pQCDcoe3}
\end{eqnarray}
Analytic expressions for $\eta_{ij}^{(0)}$, $\eta_{ij}^{(1)}$ and $\eta_{ij}^{(2)}$ can be obtained up to NNLO level from Refs.\cite{Anastasiou:2002yz, Ravindran:2003um}. It is important to note that there are two types of large logarithmic terms $\ln(\mu_r/m_{H})$ and $\ln(\mu_r/m_{t})$ in the NNLO coefficient $\eta_{gg}^{(2)}(s,M,R)$. Thus a single guessed scale, using conventional scale-setting, such as $\mu_r=m_H$, cannot eliminate all of the large logarithmic terms. In contrast, PMC scale-setting deals with such problems, as well as setting multiple renormalization scales for physical applications which depend on multiple kinematic variables. Examples have been presented for heavy-quark pair production via $q\bar{q}$ fusion~\cite{Brodsky:1995ds} and hadronic $Z$ decays~\cite{Wang:2014aqa}. For example, the process $q \bar q \to Q\bar Q$ near the heavy quark ($Q$) threshold involves not only the invariant variable $\hat s \sim 4 M^2_Q$, but also the variable ${ v_{\rm rel}}^2 \hat s$ which enters the Sudakov final-state corrections.

By using the pattern determined by the $R_\delta$-scheme method~\cite{Mojaza:2012mf, Brodsky:2013vpa}, we can rewrite the $n_f$-series at each order into a corresponding $\beta$ term series:
\begin{widetext}
\begin{eqnarray}
\hat\sigma_{ij}(s,M,R)=\frac{\pi}{576 v^2} \left[r^{ij}_{1,0}a_s^2(\mu_r)+\left(r^{ij}_{2,0} +2\beta_0r^{ij}_{2,1}\right)a_s^3(\mu_r) +\left(r^{ij}_{3,0}+2\beta_1r^{ij}_{2,1}+3\beta_0r^{ij}_{3,1} +3\beta^2_0r^{ij}_{3,2}\right)a_s^4(\mu_r)+{\cal O}(a_s^5) \right], \label{eq:partonicbeta}
\end{eqnarray}
\end{widetext}
where $\beta_{0} = 11-{2\over 3}n_{f}$, $\beta_{1} = 102-{38\over 3} n_{f}$, and the coefficients $r^{ij}_{m,n}$ are related to the $c^{ij}_{m,n}$ as
\begin{eqnarray}
&&r^{ij}_{1,0}=c^{ij}_{1,0}, \\
&&r^{ij}_{2,0}=\frac{1}{2}\left(2c^{ij}_{2,0} + 33 c^{ij}_{2,1}\right), ~r^{ij}_{2,1}=-\frac{3 c^{ij}_{2,1}}{4}, \\
&&r^{ij}_{3,0}=\frac{1}{4}\left(-642 c^{ij}_{2,1} + 4 c^{ij}_{3,0} + 66 c^{ij}_{3,1} + 1089 c^{ij}_{3,2}\right), \nonumber\\
&&r^{ij}_{3,1}=\frac{1}{2} \left(19 c^{ij}_{2,1} - c^{ij}_{3,1} - 33 c^{ij}_{3,2}\right), ~r^{ij}_{3,2}=\frac{3 c^{ij}_{3,2}}{4} .
\end{eqnarray}
The $r^{ij}_{m,0}$ with $m$=(1,2,3) are scheme-independent conformal coefficients, whereas the $r^{ij}_{m,n}$ with $1\leq n\leq m\leq 3$ are the scheme-dependent non-conformal coefficients which determine the PMC scales at each order and are absorbed into the running coupling via the RG-equation.

Following the standard PMC scale-setting procedure, we obtain the scheme-independent conformal series for $\hat\sigma_{ij}(s,M,R)$, i.e.,
\begin{eqnarray}
\hat\sigma_{ij}(s,M,R)&=&\frac{\pi}{576 v^2} \left[r^{ij}_{1,0}a_s^2(Q^{ij}_1)+r^{ij}_{2,0}a_s^3(Q^{ij}_2)\right. \nonumber\\
&&\quad\quad\quad \left. +r^{ij}_{3,0}a_s^4(Q^{ij}_3)+{\cal O}(a_s^5) \right],
\label{eq:partonicapmc}
\end{eqnarray}
where the $Q^{ij}_{m}$ with $m=(1,2,3)$ stand for the LO, NLO and NNLO PMC scales, respectively.

As we have emphasized, the renormalization scales and the resulting effective number of flavors $n_f$ obtained in pQCD by shifting the $\beta$ terms into their respective running couplings are in general distinct at each order, reflecting different virtualities of the skeleton graphs of the subprocesses as a function of phase-space. More explicitly, as indicated by Eq.(\ref{eq:partonicbeta}), new types of $\{\beta_i\}$-terms appear at each order, indicating different $\alpha_s$-running behaviors at different perturbative orders; this also shows the importance of identifying different renormalization scales at each order. Furthermore, the PMC scales in the resulting perturbative series properly absorb all of the non-conformal $\{\beta_i\}$-terms via the RG-equation, thus determining the correct arguments of the strong couplings at each order; they are given by
\begin{eqnarray}
\ln\frac{(Q^{ij}_1)^2}{\mu_r^2}&=& -\frac{r^{ij}_{2,1}}{r^{ij}_{1,0}} \nonumber\\
&& + \frac{3 ((r^{ij}_{2,1})^2 - r^{ij}_{1,0} r^{ij}_{3,2})\beta_0}{2 (r^{ij}_{1,0})^2}a_s(\mu_r)\nonumber\\&&+{\cal O}(a_s^2), \label{higgsPMCscale1}
\\
\ln\frac{(Q^{ij}_2)^2}{\mu_r^2}&=& -\frac{r^{ij}_{3,1}} {r^{ij}_{2,0}}+{\cal O}(a_s).
 \label{higgsPMCscale2}
\end{eqnarray}

A small residual dependence on the initial scale can appear in PMC predictions due to the perturbative nature of the PMC scales; this is caused by the uncomputed NNNLO terms and the higher order $\{\beta_{i}\}$-terms. Since there are no $\{\beta_{i}\}$-terms available to set the NNLO PMC scale $Q^{ij}_{3}$, we will, as a rule, set its value to the same value as the last known PMC scale $Q^{ij}_{2}$. In contrast to the conventional scale-setting procedure where a single renormalization scale is guessed, the PMC scales are fixed via the RG-equation, and the resulting residual scale-dependence is effectively independent of the initial choice of scale~\cite{Wu:2013ei}. After PMC scale setting, the remaining conformal series with $\beta_i=0$ is scheme-independent. The scheme-and-scale ambiguities are thus effectively eliminated by applying the PMC.

The PMC formulas given above are applicable to all Higgs hadroproduction channels. The production channels with $(ij) = (gg)$, $(q{\bar q})$, $(gq)$, $(g\bar{q})$, $(qq')$ and $(\bar{q}\bar{q}')$ are distinct and non-interfering, and thus their PMC scales can be determined independently. As indicated by Eqs.(\ref{pQCDcoe1}, \ref{pQCDcoe2}, \ref{pQCDcoe3}), the different channels contribute to the hadronic Higgs production cross section at different orders. Thus, up to NNLO level, we can fix two PMC scales for the $(gg)$-channel and one PMC scale for the $(q{\bar q})$-, $(gq)$- and $(g\bar{q})$-channels. In the case of the $(qq')$- and $(\bar{q}\bar{q}')$-channels, there are no higher-order computations available which can be used to set their PMC scales; thus we will set their values as the initial scale $\mu_r$. Because the $(qq')$- and $(\bar{q}\bar{q}')$-channels are negligible in comparison to the dominant $(gg)$-channel, their scale uncertainties will not appreciably affect our final results.

\section{Numerical results and discussions} \label{sect3}

For our numerical computations, we will take the Higgs mass $m_H$=125 GeV, the top-quark pole mass $m_t=173.3$ GeV~\cite{toppole}, and assume the CT10 set of PDFS obtained by the CTEQ group~\cite{Lai:2010vv}. The running behavior of $\alpha_s(Q^2) $ at NNLO order is adopted, and its magnitude is determined taking the value $\alpha_s(M_Z)=0.118$ used for CT10.

\subsection{General properties of Higgs boson hadroproduction}

\begin{table}[htb]
\centering
\begin{tabular}{|c|c|c|c|c|c|c|c|c|}
\hline
& \multicolumn{4}{c|}{Conventional} & \multicolumn{4}{c|}{PMC } \\
\hline
($ij$) & LO & NLO & N$^2$LO & Total & LO & NLO & N$^2$LO & Total \\
\hline
($gg$) & 6.02 & 7.53 & 5.21 & 18.76 & 6.02 & 9.58 & 8.01 & 23.61 \\
($gq$) & 0.00 & -0.11 & -0.31 & -0.42 & 0.00 & -0.32 & 0.21 & -0.11 \\
($g\bar{q}$) & 0.00 & -0.08 & -0.16 & -0.24 & 0.00 & -0.17 & 0.02 & -0.15 \\
($q\bar{q}$) & 0.00 & 0.008 & 0.006 & 0.014 & 0.00 & 0.007 & 0.007 & 0.014 \\
($qq+qq'$) & 0.00 & 0.00 & 0.006 & 0.006 & 0.00 & 0.00 & 0.006 & 0.006 \\
($\bar{q}\bar{q}+\bar{q}\bar{q}'$) & 0.00 & 0.00 & 0.001 & 0.001 & 0.00 & 0.00 & 0.001 & 0.001 \\
\hline
\end{tabular}
\caption{The cross-sections $\sigma^{(ij)}_{m}$ (in units of pb) for Higgs boson hadroproduction by applying conventional and PMC scale-settings at the LHC with $\sqrt{S}=8$ TeV, where $m=$LO, NLO and NNLO, respectively. The initial choice of renormalization scale and the factorization scale are taken as $\mu_r=\mu_f=m_H$. } \label{Higgspro}
\end{table}

We will first compare the pQCD predictions up to NNLO level for the total Higgs boson production cross-sections at the LHC at $\sqrt{S}=8$ TeV by applying conventional and PMC scale-settings, respectively. The results are presented in Table~\ref{Higgspro}, where the cross-sections for individual production channels are presented and `Total' stands for the sum of the cross-sections up to NNLO level; i.e., $\sigma^{(ij)}_{\rm Total}=\sum^{\rm NNLO}\limits_{m=\rm LO} \sigma^{(ij)}_{m}$. For uniformity, the initial renormalization scale and the factorization scale are each taken as $\mu_r=\mu_f=m_H$. Table~\ref{Higgspro} shows that the gluon-fusion channel provides more than $95\%$ of the contribution to the Higgs hadroproduction cross-section, whereas the remaining channels provide small or even negligible contributions. (The negative values of the $gq$ and $g\bar{q}$ contributions reflect the fact that they are cross-term contributions.)

\begin{figure}[tb]
\centering
\includegraphics[width=0.5\textwidth]{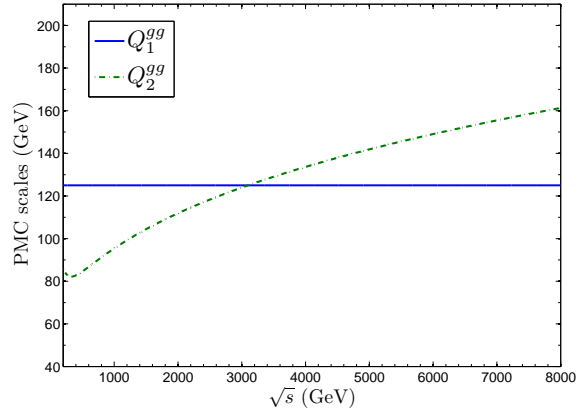}
\caption{Two PMC scales versus the subprocess center-of-mass energy $\sqrt{s}$ for the dominant $(gg)$-channel. $m_H=125$ GeV and $\mu_r=\mu_f=m_H$. }
\label{PMCscalesqrt}
\end{figure}

Table~\ref{Higgspro} shows that before and after applying the PMC, the pQCD convergence behaves quite differently for the different production channels. In the case of the $(gq)$-channel and $(g\bar{q})$-channels, the improvement is obvious: Their NNLO contributions using conventional scale-setting are even larger than the NLO contribution; in contrast, the NNLO contributions become smaller after applying the PMC. This indicates that the conformal terms in the $(gq)$-channel or $(g\bar{q})$-channel satisfy the usual requirements of pQCD convergence.

In the case of the dominant $(gg)$-channel, the pQCD convergence does not show explicit improvement at the NNLO level, even after applying the PMC -- in contrast to our previous PMC applications. This can be explained by the properties of the two PMC scales $Q^{gg}_1$ and $Q^{gg}_2$. The LO PMC scale $Q^{gg}_1$ is fixed to be $m_H$. This is due to the fact that the non-conformal coefficients $r^{gg}_{2,1}(s,M,R)$ and $r^{gg}_{3,2}(s,M,R)$ are always accompanied by the logarithmic-term $\ln(\mu^2_r/m^2_{H})$; the elimination of the nonconformal $\beta$ terms is thus equivalent to the elimination of these logarithmic terms. The LO PMC scale $Q^{gg}_1$ is fixed to be $m_H$, independent of the choice of the initial scale. In contrast, the NLO PMC scale $Q^{gg}_2$ is determined by the non-conformal coefficient $r^{gg}_{3,1}(s,M,R)$ which is a function of the subprocess center-of-mass energy ($\sqrt{s}$). To show this explicitly, we have displayed the PMC scales $Q^{gg}_1$ and $Q^{gg}_2$ versus the subprocess center-of-mass energy $\sqrt{s}$ in Fig.(\ref{PMCscalesqrt}). Fig.(\ref{PMCscalesqrt}) shows that $Q^{gg}_2$ increases with $\sqrt{s}$; this is consistent with its QED analog obtained using the GM-L scale-setting approach. The value of the scale $Q^{gg}_2$ in the threshold region is in fact smaller than $m_H$, thus yielding large cross-sections at both the NLO and the NNLO level. One thus obtains significantly larger total cross-sections in comparison with the rates predicted using conventional scale-setting.

Table~\ref{Higgspro} shows the pQCD convergence for this particular process is not improved by applying the PMC; this is caused by the presence of a large conformal term even at the NNLO order. Thus a NNNLO or higher calculation is important. A NNNLO computation for the dominant gluon-fusion channel has in fact been performed~\cite{Anastasiou:2015ema}; it has a numerically small NNNLO contribution, even using conventional scale-setting, suggesting improved pQCD convergence at this order. However, the published NNNLO calculation does not provide the $\beta$ terms needed for a PMC analysis, as required for a precise determination of the PMC scales $Q^{gg}_{2,3}$. Nevertheless, as will be shown in following subsections, we can achieve precise predictions for the Higgs boson production at the NNLO level by applying the PMC, regardless of the weak pQCD convergence of the conformal terms~\footnote{We emphasize that the purpose of PMC is to solve the renormalization scheme-and-scale ambiguities; improved QCD convergence is a natural byproduct due to the elimination of divergent renormalon terms; however, this does not affect processes with large conformal terms.}.

\subsection{An analysis of renormalization scale dependence before and after PMC scale-setting}

In the following, we will concentrate on the dominant gluon-fusion production channel. It provides as an explicit example for illustrating how the renormalization scale dependence is changed before and after PMC scale-setting.

\begin{table}[htb]
\centering
\begin{tabular}{|c|c|c|c|c|c|c|c|c|}
\hline
& \multicolumn{4}{c|}{Conventional} & \multicolumn{4}{c|}{PMC} \\
\hline
$\mu_r$ & LO & NLO & N$^2$LO & Total & LO & NLO & N$^2$LO & Total \\
\hline
$m_H/4$ & 9.42 & 10.64 & 3.50 & 23.56 & 6.02 & 9.58 & 8.01 & 23.61 \\
$m_H/2$ & 7.43 & 8.89 & 4.82 & 21.14 & 6.02 & 9.58 & 8.01 & 23.61 \\
$m_H$ & 6.02 & 7.53 & 5.21 & 18.76 & 6.02 & 9.58 & 8.01 & 23.61 \\
$2m_H$ & 4.98 & 6.45 & 5.19 & 16.62 & 6.02 & 9.58 & 8.01 & 23.61 \\
$4m_H$ & 4.19 & 5.58 & 4.95 & 14.72 & 6.02 & 9.58 & 8.01 & 23.61 \\
\hline
\end{tabular}
\caption{The gluon-fusion cross-section $\sigma^{(gg)}_{m}$ (in units of pb) using conventional and PMC scale-settings at $\sqrt{S}=8$ TeV. The results are given for five choices of initial scale, i..e $\mu_r=m_H/4$, $m_H/2$, $m_H$, $2m_H$ and $4m_H$, respectively. In addition, we take $\mu_f=m_H$. } \label{Higgsproscale}
\end{table}

We list the gluon-fusion cross-sections at $\sqrt{S}=8$ TeV using conventional and PMC scale-settings in Table \ref{Higgsproscale} for five choices of the initial scale $\mu_r=m_H/4$, $m_H/2$, $m_H$, $2m_H$, and $4m_H$. Table \ref{Higgsproscale} shows that for conventional scale-setting, $\sigma^{(gg)}_{\rm Total} =18.76^{+12.69\%}_{-11.41\%}$ pb for $\mu_r\in[m_H/2,2m_H]$, and $\sigma^{(gg)}_{\rm Total}=21.14^{+11.45\%}_{-11.26\%}$ pb for $\mu_r\in[m_H/4,m_H]$. These NNLO predictions are consistent with the NNNLO results within errors~\cite{Anastasiou:2015ema}: $\sigma^{(gg)}_{\rm Total}=18.90^{+3.08\%}_{-5.02\%}$ pb for $\mu_r\in[m_H/2,2m_H]$ and $\sigma^{(gg)}_{\rm Total}=19.47^{+0.32\%}_{-2.99\%}$ pb for $\mu_r\in[m_H/4,m_H]$. The results show that by including the NNNLO-terms, the renormalization scale uncertainty for total cross-sections can indeed be improved; e.g., if one takes $\mu_r \in [m_H/2,2m_H]$, the scale uncertainty is reduced from $24\%$ to $8\%$; and if one takes $\mu_r \in [m_H/4,m_H]$, the scale uncertainty is reduced from $23\%$ to $3\%$.

\begin{figure}[htb]
\centering
\includegraphics[width=0.50\textwidth]{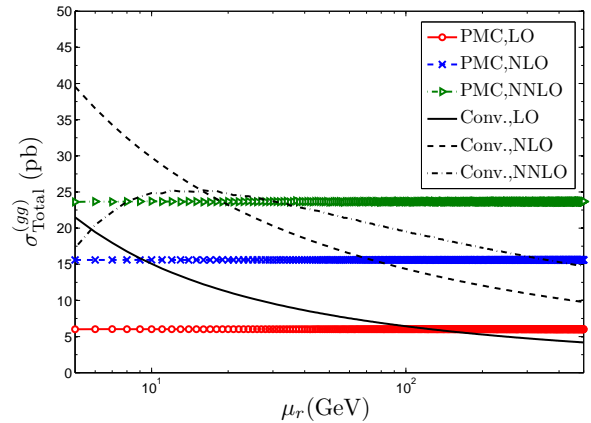}
\caption{The gluon-fusion total cross-sections $\sigma^{(gg)}_{\rm Total}$ up to LO, NLO and NNLO levels versus the initial scale $\mu_r$ by applying conventional and PMC scale-settings for CM collision energy $\sqrt{S}=8$ TeV. }
\label{HiggscsConvPMC}
\end{figure}

Fig.(\ref{HiggscsConvPMC}) shows how the pQCD prediction for the gluon-fusion total Higgs hadroproduction cross-section changes as more higher-order terms are included. The initial scale dependence of the gluon-fusion total cross-section $\sigma^{(gg)}_{\rm Total}$ is displayed at LO, NLO and NNLO levels, respectively. The figure shows that using conventional scale-setting, the LO and NLO total cross-sections $\sigma^{(gg)}_{\rm Total}|_{\rm LO}$ and $\sigma^{(gg)}_{\rm Total}|_{\rm NLO}$ depend heavily on the choice of $\mu_r$, whereas the initial scale dependence of the NNLO total cross-section $\sigma^{(gg)}_{\rm Total}|_{\rm NNLO}$ becomes smaller. The predicted value of $\sigma^{(gg)}_{\rm Total}|_{\rm NNLO}$ first increases and then decreases with increasing $\mu_r$; achieving its maximum value at $\mu_{r}\sim14$ GeV. The results are consistent with the expectation that the initial-scale sensitivity is progressively decreased by performing higher-and-higher order calculations. In contrast, after applying the PMC, the initial scale dependence is negligible even at the LO level, since the LO PMC scale is fixed to be $m_H$, independent of the choice of $\mu_r$. At the NNLO level, the resulting total cross-section is $\sigma^{(gg)}_{\rm Total}\cong 23.61$ pb for a wide range of initial scales $\mu_r$. This negligible residual dependence on the choice of the initial renormalization scale for the PMC predictions is expected, since the PMC renormalization scales at each order are precisely determined using the RG-equation.

Furthermore, by analyzing the pQCD series in detail, we find that the small renormalization scale dependence of the high-order prediction on the total cross-section $\sigma^{(gg)}_{\rm Total}$ using conventional scale-setting is caused by correlations of the scale dependence among different orders; however, the large scale dependence at each order cannot be repaired by using a guessed scale. As an explanation, we define a ratio $\kappa_m$ to measure the scale dependence of the gluon-fusion cross-section at each order,
\begin{eqnarray}
\kappa_m=\frac{\left. \sigma^{(gg)}_{m}\right|_{\mu_{r}=m_H/2} -\left. \sigma^{(gg)}_{m}\right|_{\mu_{r}=2m_H}}{\left.\sigma^{(gg)}_{m} \right|_{\mu_{r}=m_H}},
\end{eqnarray}
where the subscript $m$ stands for LO, NLO, NNLO, and Total, respectively. Applying the conventional scale-setting, we obtain
\begin{eqnarray}
\kappa_{\rm LO}=41\%,~\kappa_{\rm NLO}=32\%,~\kappa_{\rm NNLO}=-7\%. \label{seperrorconv}
\end{eqnarray}
This shows that by using conventional scale-setting, the scale dependence at each perturbative order is rather large. The weighted average of those separate scale errors gives a smaller total scale error, $\kappa_{\rm Total}=24\%$, which can be further reduced down to $\sim 8\%$~\cite{Anastasiou:2015ema} by including the NNNLO contributions.

On the other hand, Table~\ref{Higgsproscale} shows that the magnitudes of all $\kappa_m$ are negligible after applying the PMC, showing the renormalization scale uncertainties for both the separate cross-sections at each order and the total cross-section are simultaneously eliminated.

After applying the PMC, the total cross-section $\sigma^{(gg)}_{\rm Total}$ is $\sim 26\%$ larger than the one for $\mu_r=m_H$ using conventional scale-setting. As shown by Table~\ref{Higgsproscale}, if one sets $\mu_r=m_H/2$ or $\mu_r=m_H/4$, the conventional pQCD convergence is better than the case of $\mu_r=m_H$, and its total cross-section is close to the PMC prediction; i.e., within an error of $10\%$. Thus, for conventional scale-setting, the best choice of an effective renormalization scale would be around $m_H/2$ or $m_H/4$ other than the usually chosen value of $m_H$. This prescription has already been suggested in Refs.\cite{Anastasiou:2015ema, Anastasiou:2005qj, Anastasiou:2011pi}; the PMC thus provides the explanation for this choice.

\subsection{Predictions for ``uncalculated" higher-order contributions}

It is useful to estimate the magnitude of contributions from ``uncalculated" high-order perturbative terms. Each scale-setting approach has different conventions for estimating the theory uncertainty.

\begin{figure}[tb]
\centering
\includegraphics[width=0.5\textwidth]{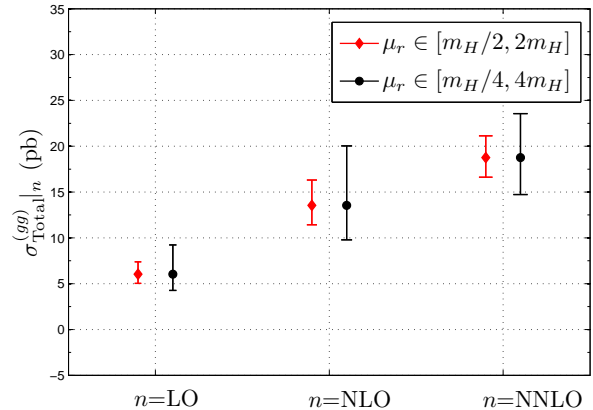}
\caption{Results for gluon-fusion total cross-section $\sigma^{(gg)}_{\rm Total}|_{n}=\sum^{n} \limits_{i=\rm {LO}} {\cal C}_i\, a_s^{i+1}$ using conventional scale-setting, where $n$ stands for LO, NLO or NNLO, respectively. The error bars represent the prediction for the magnitude of ``uncalculated" higher-order terms which are obtained by varying $\mu_r\in[m_H/2, 2m_H]$ or $\mu_r\in[m_H/4, 4m_H]$ in all ``known" lower-order terms. }
\label{errorbarConv}
\end{figure}

Under the conventional scale-setting, one usually estimates the magnitude of unknown higher-order pQCD corrections to the Higgs hadroproduction cross-section by varying the renormalization scale in the range $\mu_r \in[m_H/2, 2m_H]$. However this {\it ad hoc} procedure only gives a rough estimate of the uncalculated non-conformal $\beta$ terms at higher-orders, and it has no sensitivity to the conformal contributions at the same order which may have equal importance~\cite{Wu:2013ei}. As will be shown in the following, the Higgs hadroproduction cross-section illustrates the unreliability of conventional error estimates.

Schematically, we rewrite the gluon-fusion total cross-section using conventional scale-setting as
\begin{equation}
\sigma^{(gg)}_{\rm Total}|_{n}=\sum^{n} \limits_{i=\rm {LO}} {\cal C}_i\; a_s^{i+1}(\mu_r), \label{Convgg}
\end{equation}
where $n$ stands for LO, NLO or NNLO, respectively. The results are presented in Fig.(\ref{errorbarConv}), in which the error bars stand for the conventional estimate of ``uncalculated" higher-order contributions to the cross-sections, obtained by varying $\mu_r\in[m_H/2,2m_H]$ or $\mu_r\in[m_H/4,4m_H]$ in all of the ``known" lower-order terms. For the conventional case of $\mu_r\in[m_H/2,2m_H]$, Fig.(\ref{errorbarConv}) confirms the observation of Ref.\cite{Forte:2013mda}, which shows that the ``true" NLO cross-section $\sigma^{(gg)}_{\rm Total}|_{\rm NLO}$ together with its error bar, is outside of the predicted NLO values from the LO calculation, and the ``true" NNLO cross-section $\sigma^{(gg)}_{\rm Total}|_{\rm NNLO}$, together with its error bar, is outside of the predicted NNLO values from the NLO calculation. By taking $\mu_r\in[m_H/4,m_H]$, the condition becomes better, but the ``true" NLO cross-section $\sigma^{(gg)}_{\rm Total}|_{\rm NLO}$ together with its error bar is still outside of the predicted NLO values from the LO calculation. This shows the conventional way of estimating the magnitude of ``unknown" terms by varying $\mu_r$ with a certain range is invalid for the gluon-fusion Higgs production channel at least at the lower orders.

As shown in Table~\ref{Higgsproscale}, the conformal series derived by applying the PMC shows a weak perturbative convergence up to NNLO level, indicating that the conformal terms for this particular process are very important, at least at the NNLO level. This explains why the calculated higher-order predictions are always outside of the error bars predicted from the lower-order cross-sections using conventional error estimates -- its higher-order estimation does not take into account the large conformal contributions.

On the other hand, after applying the PMC, the situation is quite different. The PMC renormalization scales are determined unambiguously, and they are independent of the choice of the initial scale. Thus they cannot be varied; otherwise, one would explicitly break RG-invariance, leading to an unreliable prediction. Thus, the standard way of predicting unknown higher-order contributions is not applicable to PMC predictions. Instead, we adopt a more conservative practice for the error estimate for PMC predictions~\cite{Wu:2014iba}; i.e., we will define the PMC error bar to match the value of the contribution from the last known perturbative order. More explicitly, after applying the PMC, the gluon-fusion total cross-section (\ref{Convgg}) changes to
\begin{eqnarray}
\sigma^{(gg)}_{\rm Total}|_n=\sum^{n}\limits_{i=\rm {LO}} \tilde{\cal C}_i \; a_s^{i+1}(Q^{gg}_i),
\end{eqnarray}
where $\tilde{C}_i$ is the $i_{\rm th}$-order conformal coefficient, and $n$ stands for LO, NLO and NNLO, respectively. Thus the PMC estimate of the uncalculated higher-order terms for an $i_{\rm th}$-order calculation is $\pm |\tilde{\cal C}_{i} a^{i+1}_s(Q^{gg}_i)|_{\rm MAX}$, where both $\tilde{\cal C}_{i}$ and $a_s(Q^{gg}_i)$ are calculated by varying the scale $\mu_r\in[m_H/2,2m_H]$ and the symbol ``MAX'' stands for the maximum value of $|\tilde{\cal C}_{i} a^{i+1}_s(Q^{gg}_i)|$ within this region. This procedure is natural for the PMC, since after PMC scale-setting, the main uncertainty is from the last term at this order with its unfixed PMC scale. As shown by Eq.(\ref{seperrorconv}), the conventional scale errors dominate at each perturbative order; thus this approach for the error estimate cannot be applied to conventional scale-setting.

\begin{figure}[tb]
\centering
\includegraphics[width=0.5\textwidth]{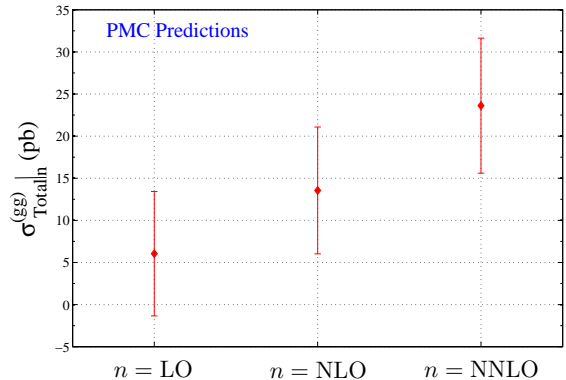}
\caption{Results for gluon-fusion total cross-section $\sigma^{(gg)}_{\rm Total}|_{n}=\sum^{n} \limits_{i=\rm {LO}} \tilde{\cal C}_i\, a_s^{i+1}(Q^{gg}_i)$ by applying PMC scale-setting; here $n$ stands for LO, NLO or NNLO, respectively. The error bar for $i_{\rm th}$-order stands for a conservative prediction of ``uncalculated" higher-order terms, which is taken as $\pm|\tilde{\cal C}_{i}\, a_s^{i+1}(Q^{gg}_i)|_{\rm MAX}$. }
\label{errorbarPMC}
\end{figure}

The conservative PMC estimates for the ``uncalculated" higher-order terms for the gluon-fusion total cross-section are displayed in Fig.(\ref{errorbarPMC}). Because of large conformal terms at the NLO and NNLO levels, the pQCD convergence cannot be greatly improved as is usual the case; thus the PMC predictions have large error bars. In contrast to the conventional error estimates obtained by varying the renormalization scale in the range $\mu_r\in[m_H/2,2m_H]$ which shown by Fig.(\ref{errorbarConv}), Fig.(\ref{errorbarPMC}) shows the ¡°true¡± values of the higher-order cross sections using the PMC obtained by defining the error bar to match the value of the contribution from the last known perturbative order are consistently within the error bars defined from the lower-order calculation.

As additional higher order terms in pQCD series become available, both the conformal terms and the PMC scales become more precisely determined, and the predictive power of PMC predictions will be improved. For example, for the observables $\Gamma(H\to b\bar{b})$ and $R(e^+e^-)$ up to four-loop level, it has been found that the predicted error bars from the `unknown' higher-order corrections quickly approach stability, and the resulting higher-order PMC predictions are well within the error bars predicted from the ``known" lower-order terms~\cite{Wu:2014iba}.

\subsection{A discussion of the factorization scale dependence}

After applying the PMC, the renormalization scale ambiguity is eliminated. One of the remaining uncertainties is from the choice of factorization scale. The determination of factorization scale is a separate issue, which may be solved by matching nonpertubative bound-state dynamics with perturbative DGLAP evolution~\cite{Gribov:1972ri, Altarelli:1977zs, Dokshitzer:1977sg}. Recently, by using the light-front holography~\cite{Brodsky:2011pw, deTeramond:2012rt}, it has been shown that the matching of high-and-low scale regimes of $\alpha_s$ can determine the scale which sets the interface between perturbative and nonperturbative hadron dynamics~\cite{Deur:2014qfa, Deur:2016cxb, Deur:2016tte}.

In the following illustration, we will adopt the usual convention of choosing the range $\mu_{f} \in[m_H/2,2m_H]$ to predict the factorization scale dependence.

As is the case for top-pair hadroproduction~\cite{Wang:2014sua}, we have found that the usual factorization scale dependence can also be largely suppressed for Higgs hadroproduction once the renormalization scale is set by the PMC. For example, in the case of the dominant gluon-fusion channel using conventional scale-setting, we obtain,
\begin{eqnarray}
{\sigma^{(gg)}_{\rm Total}}_{|7\rm TeV} &=& \left(14.46^{+0.90}_{-0.85}\right)~{\rm pb},\nonumber\\
{\sigma^{(gg)}_{\rm Total}}_{|8\rm TeV} &=& \left(18.76^{+1.23}_{-1.18}\right)~{\rm pb}, \nonumber\\
{\sigma^{(gg)}_{\rm Total}}_{|13\rm TeV} &=& \left(45.32^{+3.45}_{-3.50}\right)~{\rm pb}, \nonumber
\end{eqnarray}
where $\mu_r =m_H$, the central value is for $\mu_f=m_H$, and the errors are for $\mu_{f} \in[m_H/2,2m_H]$. In contrast, after applying the PMC scale-setting, we obtain
\begin{eqnarray}
{\sigma^{(gg)}_{\rm Total}}_{|7\rm TeV} &=& \left(18.27^{+0.33}_{-0.83}\right)~{\rm pb}, \nonumber\\
{\sigma^{(gg)}_{\rm Total}}_{|8\rm TeV} &=& \left(23.61^{+0.27}_{-0.94}\right)~{\rm pb}, \nonumber\\
{\sigma^{(gg)}_{\rm Total}}_{|13\rm TeV} &=& \left(56.48^{-0.65}_{-1.10}\right)~{\rm pb}. \nonumber
\end{eqnarray}
These results show the factorization scale dependence can be suppressed due to the correlation of the large logarithmic terms involving the renormalization and factorization scales; i.e. $\ln\mu^2_r/m^2_H$ and $\ln\mu^2_f/m^2_H$.

\begin{table}[htb]
\centering
\begin{tabular}{c|c|cccc}
\hline
 & Tevatron & \multicolumn{4}{c}{LHC } \\
\hline
$\sqrt{S}$ & 1.96 TeV & 7 TeV & 8 TeV & 13 TeV & 14 TeV \\
\hline
Conv. & $0.63^{+0.13}_{-0.11}$ & $13.92^{+2.25}_{-2.06}$ & $18.12^{+2.87}_{-2.66}$ & $44.26^{+6.61}_{-6.43}$ & $50.33^{+7.47}_{-7.31}$ \\
PMC  & $0.86^{+0.13}_{-0.12}$ & $18.04^{+1.36}_{-1.32}$ & $23.37^{+1.65}_{-1.59}$ & $56.34^{+3.45}_{-3.00}$ & $63.94^{+3.88}_{-3.30}$ \\
\hline
\end{tabular}
\caption{Scale uncertainties for $\sigma_{\rm ggH}$ (in units of pb) using the conventional (Conv.) versus PMC scale-settings, obtained by varying $\mu_r \in[m_H/2,2m_H]$ and $\mu_f \in[m_H/2,2m_H]$. Here $\sigma_{\rm ggH}$ stands for the sum of the cross-sections $\sigma^{(ij)}_{\rm Total}$ with $(ij) = (gg)$, $(q{\bar q})$, $(gq)$, $(g\bar{q})$, $(qq')$, respectively.} \label{HiggsproTlhc}
\end{table}

As a summary, we present total hadronic cross-section $\sigma_{\rm ggH}$ for the Tevatron and LHC in Table~\ref{HiggsproTlhc}, where the errors are for $\mu_r \in[m_H/2,2m_H]$ and $\mu_f \in[m_H/2,2m_H]$. For convenience, we let $\sigma_{\rm ggH}$ stand for the sum of the cross-sections $\sigma^{(ij)}_{\rm Total}$ with $(ij) = (gg)$, $(q{\bar q})$, $(gq)$, $(g\bar{q})$ and $(qq')$, respectively. The errors are the squared averages of the ones for all the hadronic production channels. To compare with the total hadronic cross-sections using conventional scale-setting, the central values of the PMC ones are increased by $\sim37\%$ at the Tevatron, and by $\sim30\%$ at the LHC for $\sqrt{S}=$7, 8, 13 and 14 TeV, respectively. After applying the PMC, the main uncertainty is from the choice of factorization scale which is generally smaller than the uncertainty using conventional renormalization scale setting.

\subsection{An estimate of the total inclusive cross-section for Higgs production at the LHC}

In order to compare our PMC predictions with recent LHC measurements on the Higgs boson production cross-section~\cite{Aad:2015lha, TOTCS:ATLAS}, one must, in addition to the hadronic channel ($\sigma_{\rm ggH}$), include contributions from other known production modes, such as the vector-boson fusion production process, the $WH/ZH$ Higgs associated production process, the associated Higgs production with heavy quarks, etc. We will let $\sigma_{\rm xH}$ stand for the sum of those cross-sections. Here $x$ stands for $Z$+$W$+$t\bar{t}$+$b\bar{b}$+$\cdots$, and $\sigma_{\rm EW}$ stands for the electroweak correction; this is necessary for practical comparisons with data.

The prediction for the total inclusive cross-section ($\sigma_{\rm Incl}$) for $pp\to HX$ production is given by $\sigma_{\rm ggH}+\sigma_{\rm xH}+\sigma_{\rm EW}$. The value of $\sigma_{\rm xH}$ and $\sigma_{\rm EW}$ are small in comparison to the dominant $\sigma_{\rm ggH}$ contribution. For example, taking $\sqrt{S}=8$ TeV and $m_H=125$ GeV, one predicts $\sigma_{xH}=3.08+0.10$ pb~\cite{Heinemeyer:2013tqa, Aad:2015lha}; the electro-weak correction up to two-loop level only leads to a $+5.1\%$ shift with respect to the NNLO QCD cross sections~\cite{Actis:2008ts, Actis:2008ug}. Thus, we directly adopt their values using conventional scale-setting which have been summarized in Refs.\cite{Heinemeyer:2013tqa, Aad:2015lha, Actis:2008ts, Actis:2008ug} to do our prediction.

\begin{table}[htb]
\centering
\begin{tabular}{c|ccc}
\hline
 $\sigma_{\rm Incl}$ & 7 TeV & 8 TeV & 13 TeV \\
\hline
ATLAS($H\to\gamma\gamma$)~\cite{Aad:2015lha, TOTCS:ATLAS} & $35^{+13}_{-12}$ & $30.5^{+7.5}_{-7.4}$ & $40^{+31}_{-28}$ \\
ATLAS($H\to ZZ^{*}\to 4l$)~\cite{Aad:2015lha, TOTCS:ATLAS} & $33^{+21}_{-16}$ & $37^{+9}_{-8}$ & $12^{+25}_{-16}$ \\
LHC-XS~\cite{Heinemeyer:2013tqa} & $17.5\pm1.6$ & $22.3\pm2.0$ & $50.9^{+4.5}_{-4.4}$ \\
PMC predictions & $21.21^{+1.36}_{-1.32}$ & $27.37^{+1.65}_{-1.59}$ & $65.72^{+3.46}_{-3.01}$ \\
\hline
\end{tabular}
\caption{Total inclusive cross-sections (in units of pb) for Higgs production at the LHC for CM collision energies $\sqrt{S}=7$, 8 and 13 TeV, respectively. The inclusive cross-section is defined as the sum $\sigma_{\rm Incl}=\sigma_{\rm ggH}+\sigma_{\rm xH}+\sigma_{\rm EW}$. } \label{HiggsproATLAS}
\end{table}

\begin{figure}[htb]
\centering
\includegraphics[width=0.5\textwidth]{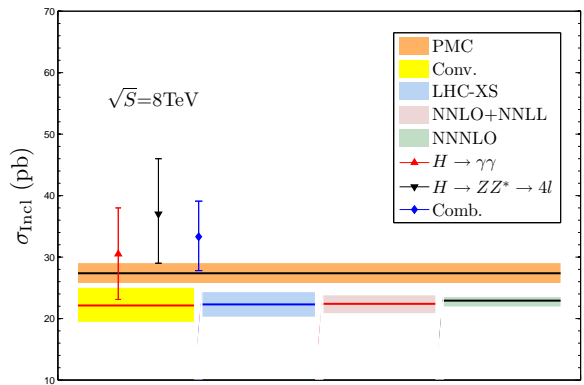}
\caption{Comparison of the NNLO conventional versus PMC predictions for the total inclusive cross-section $\sigma_{\rm Incl}$ with the latest ATLAS measurements at 8TeV~\cite{Aad:2015lha}. The LHC-XS predictions~\cite{Heinemeyer:2013tqa}, the NNLO+NNLL prediction~\cite{deFlorian:2012yg}, and the NNNLO prediction~\cite{Anastasiou:2015ema} are presented as a comparison. The solid lines are central values. }
\label{ATLASConvPMC}
\end{figure}

We present the predicted PMC total inclusive cross-section $\sigma_{\rm Incl}$ at the LHC for several CM collision energies in Table~\ref{HiggsproATLAS}, in comparison with recent LHC ATLAS measurements~\cite{Aad:2015lha, TOTCS:ATLAS}, for $H\to\gamma\gamma$ and $H\to ZZ^{*}\to 4l$ decay channels. The SM results predicted by LHC-XS group~\cite{Heinemeyer:2013tqa} is shown as a comparison. The inclusive cross-section increases with increasing hadron-hadron collision energy. To compare with the central LHC-XS predictions~\cite{Heinemeyer:2013tqa}, our PMC results are increased by about $21\%$, $23\%$ and $29\%$ for $\sqrt{S}=7$, 8 and 13 TeV, respectively. Because of the large uncertainty for the ATLAS data, we need more data to draw definite conclusion on the SM predictions \footnote{A recent ATLAS measurement for $\sqrt{S}=13$ TeV gives~\cite{TOTCSrecent:ATLAS}, $\sigma_{\rm Incl}=59.0^{+9.7}_{-9.2}(\rm stat.)^{+4.4}_{-3.5}(\rm syst.)$ pb, which shows a better agreement with our present PMC prediction.}. The more accurate measurements with high integrated luminosity for $\sqrt{S}$=13 TeV shall be helpful to test the PMC and conventional predictions. For the case of the ATLAS data at 8 TeV~\cite{Aad:2015lha} which has relatively less experimental uncertainties, the PMC prediction show a much better agreement with the data. This is clearly shown by Fig.(\ref{ATLASConvPMC}), in which a comparison of our present NNLO conventional and PMC predictions for $\sigma_{\rm Incl}$ with the ATLAS measurements at 8TeV is presented.

\subsection{Estimates of the fiducial cross section $\sigma_{\rm fid}(pp\to H\to \gamma\gamma)$}

The ATLAS group has measured the ``fiducial cross section" $(\sigma_{\rm fid})$ for the process $pp\to H\to \gamma\gamma$ at different collision energies~\cite{sigmafid:ATLAS}. The measurement utilizes integrated luminosity $4.5 {\rm fb}^{-1}$ for $\sqrt{S}=7$ TeV, $20.3 {\rm fb}^{-1}$ for $\sqrt{S}=8$ TeV, and $3.2 {\rm fb}^{-1}$ for $\sqrt{S}=13$ TeV, The fiducial cross-section $\sigma_{\rm fid}$ can be written as
\begin{eqnarray}
\sigma_{\rm fid}(pp\to H\to \gamma\gamma)=\sigma_{\rm Incl}{\cal B}_{H\to \gamma\gamma}{\cal A},
\end{eqnarray}
where ${\cal A}$ is the acceptance factor, whose values for different collision energies are~\cite{sigmafid:ATLAS}, ${\cal A}|_{\rm 7TeV}=0.620\pm0.007$, ${\cal A}|_{\rm 8TeV}=0.611\pm0.012$ and ${\cal A}|_{\rm 13TeV}=0.570\pm0.006$. The ${\cal B}_{H\to \gamma\gamma}$ is the branching ratio of $H\to \gamma\gamma$. By using the $\Gamma(H\to \gamma\gamma)$ with conventional scale-setting, the LHC-XS group predicts ${\cal B}_{H\to \gamma \gamma} =0.00228\pm0.00011$~\cite{Heinemeyer:2013tqa}. A detailed PMC analysis for $\Gamma(H\to \gamma\gamma)$ up to three-loop levels have been given in Ref.\cite{Wang:2013akk}. Using the formulae given there, we obtain $\Gamma(H\to \gamma\gamma)|_{\rm PMC}=9.34\times10^{-3}$ MeV for $m_H=125$ GeV. Using this value, together with Higgs total decay width $\Gamma_{\rm Total}=(4.07\pm0.16)\times 10^{-3}$ GeV~\cite{Heinemeyer:2013tqa}, we find ${\cal B}_{H\to \gamma\gamma}|_{\rm PMC}=0.00229\pm0.00009$.

Thus the main differences together with their theoretical errors for the fiducial cross-section $\sigma_{\rm fid}$ is due to the different predictions for the inclusive cross-section $\sigma_{\rm Incl}$ mentioned in the last subsection.

\begin{table}[htb]
\centering
\begin{tabular}{cccc}
\hline
$\sigma_{\rm fid}(pp\rightarrow H\rightarrow\gamma\gamma)$ & 7 TeV & 8 TeV & 13 TeV \\
\hline
ATLAS data~\cite{sigmafid:ATLAS} & $49\pm18$ & $42.5^{+10.3}_{-10.2}$ & $52^{+40}_{-37}$ \\
LHC-XS~\cite{Heinemeyer:2013tqa} & $24.7\pm2.6$ & $31.0\pm3.2$ & $66.1^{+6.8}_{-6.6}$ \\
PMC prediction & $30.1^{+2.3}_{-2.2}$ & $38.3^{+2.9}_{-2.8}$ & $85.8^{+5.7}_{-5.3}$ \\
\hline
\end{tabular}
\caption{The fiducial cross section $\sigma_{\rm fid}(pp\to H\to \gamma\gamma)$ (in units of fb) at the LHC for CM collision energies $\sqrt{S}=$7, 8 and 13 TeV, respectively. } \label{sigmafidATLAS}
\end{table}

\begin{figure}[htb]
\centering
\includegraphics[width=0.5\textwidth]{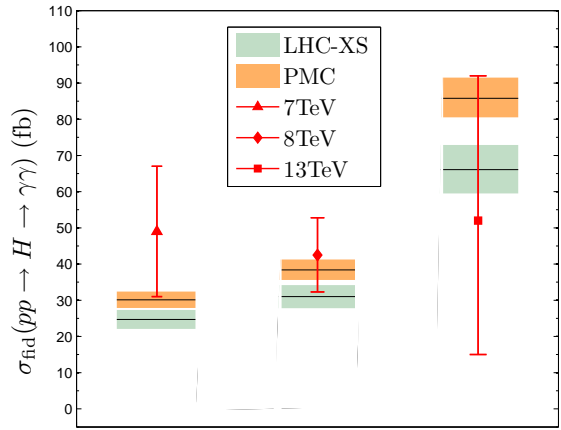}
\caption{Comparison of the PMC predictions for the fiducial cross section $\sigma_{\rm fid}(pp\to H\to \gamma\gamma)$ with ATLAS measurements at various collision energies~\cite{sigmafid:ATLAS}. The LHC-XS predictions~\cite{Heinemeyer:2013tqa} are presented as a comparison. }
\label{sigmafidATLAS7813}
\end{figure}

The PMC predictions for the fiducial cross section $\sigma_{\rm fid}(pp\to H\to \gamma\gamma)$ at the LHC for CM collision energies $\sqrt{S}=$7 TeV, 8 TeV and 13 TeV are shown in Table~\ref{sigmafidATLAS} and compared with ATLAS measurements~\cite{sigmafid:ATLAS} and the LHC-XS predictions~\cite{Heinemeyer:2013tqa} are presented. The PMC fiducial cross-sections are larger than the LHC-XS ones by $\sim22\%$, $\sim24\%$ and $\sim30\%$ for $\sqrt{S}=$7 TeV, 8 TeV and 13 TeV, respectively. Table~\ref{sigmafidATLAS} shows no significant differences between the measured fiducial cross sections and the SM predictions within the current experimental uncertainties. However, a better agreement of PMC predictions with the measurements at $\sqrt{S}=7$ TeV and 8 TeV can be obtained. This performance can be clearly shown in Fig.(\ref{sigmafidATLAS7813}), which presents the comparison of PMC predictions for $\sigma_{\rm fid}(pp\to H\to \gamma\gamma)$ with the ATLAS measurements for various CM collision energies. The forthcoming more precise measurements with higher integrated luminosity at the LHC will be important for testing the PMC theoretical predictions.

\section{Summary}
\label{sect4}

We have predicted the Higgs boson hadroproduction cross-section using PMC scale-setting. The PMC provides a systematic, rigorous way to set the renormalization scales for high-energy process at each order of perturbation theory. The PMC satisfies renormalization group invariance; i.e., PMC predictions do not depend on the choice of renormalization scheme used to regulate ultraviolet divergences and have minimal sensitivity to the initial scale choice. The PMC reduces in the $N_C \to 0$ Abelian limit to the standard Gell-Mann Low scale-setting method used for precision predictions in quantum electrodynamics. Thus the PMC treats the renormalization of all three field-theoretic components of the Standard Model and Grand Unification consistently.

After applying PMC scale-setting, the large renormalization scale and scheme uncertainties for the Higgs total and individual hadroproduction cross-sections are simultaneously eliminated. As an example, Table \ref{Higgsproscale} shows that if one uses conventional scale-setting, the predicted NNLO pQCD cross section from the dominant gluon-fusion channel at $\sqrt S = 8$ TeV varies as $\sigma^{(gg)}_{\rm Total} =18.76^{+12.69\%}_{-11.41\%}$ pb for the range of renormalization scale choices $\mu_r\in[m_H/2,2m_H]$, and has the uncertainty $\sigma^{(gg)}_{\rm Total}=21.14^{+11.45\%}_{-11.26\%}$ pb for the wider range $\mu_r\in[m_H/4,m_H]$. In contrast, after applying the PMC, we obtain the NNLO prediction $\sigma^{(gg)}_{\rm Total}\cong 23.61$ pb for $\mu_r \in[m_H/4,2m_H]$. The independence of PMC predictions on the choice of the initial renormalization scale is a feature of the PMC since the renormalization scales of the running QCD coupling at each order of perturbation theory are fixed, consistent with its renormalization group equation.

By combining the relevant Higgs boson production modes and taking the electroweak corrections into consideration, reliable pQCD scheme- and scale-independent predictions for the inclusive $pp\rightarrow H$ production cross-sections can be obtained by using the PMC. The predicted inclusive cross-section increases with increasing hadron collision energy. In comparison with the LHC-XS predictions using the guessed scale $\mu_r=m_H$, the scale-fixed PMC predictions are increased by about $21\%$, $23\%$ and $29\%$ for $\sqrt{S}=$7 TeV, 8 TeV and 13 TeV, respectively, showing good agreement with the latest LHC ATLAS measurements, especially for measurements at $\sqrt{S}=$7 TeV and 8 TeV.

We also predict the fiducial cross section $\sigma_{\rm fid}(pp\to H\to\gamma\gamma)$: The SM predictions using the PMC given in Table~\ref{sigmafidATLAS} are observed to agree with measurements within the current experimental uncertainties; the agreement of PMC predictions with the measurements at $\sqrt{S}=7$ TeV and 8 TeV is especially clear. The results demonstrate that the PMC eliminates a major theoretical uncertainty for pQCD predictions, thus increasing the sensitivity of the LHC to possible new physics beyond the SM.

\hspace{1cm}

{\bf Acknowledgements}: We thank Michael Peskin for helpful discussions on how to characterize the uncertainty of PMC predictions. This work was supported in part by the Natural Science Foundation of China under Grant No.11547010, No.11547305 and No.11275280, the Department of Energy Contract No.DE-AC02-76SF00515, and by Fundamental Research Funds for the Central Universities under Grant No.CDJZR305513. SLAC-PUB-16521. NORDITA-2016-32.

\end{document}